\documentclass[twocolumn,prd,superscriptaddress,preprintnumbers,nofootinbib]{revtex4}
\usepackage{graphicx}
\usepackage{epsfig}
\usepackage{lipsum}
\usepackage{enumitem}
\usepackage{bm}
\usepackage{latexsym,amssymb,amsmath,amsfonts,amssymb,txfonts,pxfonts,wasysym,float}
\usepackage{color}
\newcommand{\beq}[1]{\begin{equation}\label{#1}}
\newcommand{\eeq}{\end{equation}}
\newcommand{\bea}[1]{\begin{eqnarray} \label{#1}}
\newcommand{\eea}{\end{eqnarray}}
\newcommand{\ba}{\begin{array}}
\newcommand{\ea}{\end{array}}

\def\be{\begin{equation}}
\def\ee{\end{equation}}
\def\gs{\mathrel{
   \rlap{\raise 0.511ex \hbox{$>$}}{\lower 0.511ex \hbox{$\sim$}}}}
\def\ls{\mathrel{
   \rlap{\raise 0.511ex \hbox{$<$}}{\lower 0.511ex \hbox{$\sim$}}}}

\newcommand{\comment}[1]{}

\usepackage[usenames,dvipsnames]{xcolor}
\definecolor{orange}{cmyk}{0,0.5,1,0}
\definecolor{rossoCP3}{cmyk}{0,.88,.77,.40}
\definecolor{graa}{rgb}{0.8,0.8,0.8}
\definecolor{blaa}{rgb}{0.2,0.2,0.6}

\begin{document}

\title{\color{rossoCP3}{Introducing a special collection of papers in the Journal of High Energy Astrophysics on the
Early Results of China's 1st X-ray Astronomy Satellite: \textit{Insight}-HXMT
}}

\author{Diego F. Torres}

\affiliation{Instituci\'o Catalana de Recerca i Estudis Avan\c{c}ats (ICREA), E-08010 Barcelona, Spain
}

\affiliation{Institute of Space Sciences (ICE, CSIC), Campus UAB, Carrer de Magrans s/n, 08193 Barcelona, Spain
}

\affiliation{Institut d'Estudis Espacials de Catalunya (IEEC), 08034 Barcelona, Spain
}

\author{Shuang-Nan Zhang}

\affiliation{Key Laboratory of Particle Astrophysics, Institute of High Energy Physics, Chinese Academy of Sciences, 19B Yuquan Road, Beijing 100049, China
}
\affiliation{University of Chinese Academy of Sciences, Chinese Academy of Sciences, Beijing 100049, China
}
\affiliation{Key Laboratory of Space Astronomy and Technology, National Astronomical Observatories, Chinese Academy of Sciences, Beijing 100012, China
}

\maketitle

{\it Insight}-HXMT is the first Chinese X-ray astronomical mission, launched successfully on June 15, 2017, from China's Jiuquan Satellite Launch Center.
{\it Insight}-HXMT was designed to have a broad energy coverage in X-rays, from 1--250 keV, with excellent timing and adequate energy resolution at soft X-rays, and the largest effective area at hard X-rays.
This allows, in particular, to observe bright sources like X-ray binaries (XRBs) in their bright/outburst states with high cadence and high statistics at hard and soft X-rays at once. It was then expected that the {\it Insight}-HXMT mission will bring us new insights regarding the characteristics of several source classes.
Examples include characterizing High-Mass X-ray Binaries (HMXBs) and the outburst evolution of Low-Mass X-ray Binaries (LMXBs).
For instance, in HMXB systems, the region around their Alfven radius that is responsible to determine whether accretion or propeller occurs, or the region around the NS magnetic pole where the intense X-rays are supposed to be produced are specially appealing targets of study.
For the LMXBs, the evolution the outburst and the properties of the compact objects themselves are obvious priority targets for {\it Insight}-HXMT.
Due to the broad coverage in energy, {\it Insight}-HXMT is specially well suited to study the
influence of thermonuclear (type-I) X-ray bursts upon the surrounding environment.

Observations from {\it Insight}-HXMT are producing state-of-the-art samples in a broad energy band, covering both the timing and energetic domains.
Besides, so far the largest effective area at MeV, i.e, in soft gamma-rays, of the anti-coincidence detector CsI of {\it Insight}-HXMT, provides an almost unique power in monitoring flare events like Gamma-Ray Bursts (GRBs), especially short/hard GRBs suggested to be generated in binary neutron star mergers.
We are proud to present a collection of papers of the Journal of High Energy Astrophysics on the
Early Results of China's 1st X-ray Astronomy Satellite {\it Insight}-HXMT.

These papers cover the in-orbit performance, the background model, and all calibration results, which will serve as primary references of future observations and data analysis with {\it Insight}-HXMT, expected to operate in space for several more years to come.
Then, the special issue contains details of instrument calibration, both in the ground and in-flight \cite{XuFang,Luo,Liu,Li}, the
in-orbit calibration of the point-spread function \cite{Nang}
and of the background model for the low, medium, and high-energy instruments \cite{Liao,Liao2,Guo}.
The collection also gives details about the deadtime calculation method of {\it Insight}-HXMT \cite{Xiao}.
This is complicated in {\it Insight}-HXMT; due to the properties of the high-energy detector,
it is impossible to directly obtain the deadtime for a time interval less than 1 second.
This paper is complemented with studies on the time response distribution of {\it Insight}-HXMT low energy detector
\cite{Zhao2}.

Together with the technical aspects, some first results are also covered in this special collection.
One of those are the {\it Insight}-HXMT observation on 4U 1608--52 \cite{Chen}.
In this paper, Chen et al. reported on a photospheric radius expansion burst, finding that the burst and enhanced persistent emissions sum up to exceed Eddington luminosity by 40\%, speculating that the emission comes from beyond the photospheric radius, or through the Comptonization of the corona

{\it Insight}-HXMT observations on the 2019 outburst from the X-ray pulsar 4U 1901+03 are also reported in a paper of this collection\cite{Tuo}. This paper showed that the
pulse profiles significantly evolve during the outburst. The existence of two types of the profile's pattern was proposed
to be the result of a transition from a super- to a sub-critical accretion regime, and used to estimate a (model-dependent) distance to the source.

Another paper reports on Her X-1. This is the first source in which a cyclotron resonant scattering feature (CRSF) was detected. {\it Insight}-HXMT realized joint observations with NuSTAR \cite{Xiao3}, confirming that the previously observed long decay phase has ended and that the line energy keeps constant around 37.5 keV after flux correction.
Sco X-1 and Aql X-1, were also observed by HXMT early on the mission and reports on them are also to be found in this collection \cite{Jia,Gungor2}.
These sources have been extensively studied for years. However, the dependence of the properties for different types of Quasi-Periodic Oscillations (QPOs) on energy remain not fully understood, partly because of the lack of broadband simultaneous coverage in X-rays. For instance, for Sco X-1,
due to the limited effective area of the RXTE in the hard X-ray band and the relatively soft spectrum of the source, it was difficult to extend the QPO studies of this source to energies higher than 30 keV.
{\it Insight}-HXMT has now reported a complete Z-track hardness-intensity diagram and the first detection of kilo-Hz QPOs above 20 keV.
Its results indicate that the origin of all types of QPOs is non-thermal, probably at the innermost region of the accretion disk
and that the corona is geometrically inhomogeneous.
{\it Insight}-HXMT also show the need of a comptonization component in the spectrum of Aql X-1, in what appears to be a process of
partial accretion in the weak propeller stage in the 2018 outburst

Regarding black holes, first results of {\it Insight}-HXMT include constraining the spin parameter of the Cyg X-1 system, confirming it as a extreme Kerr black hole \cite{Gou}.
An outburst reported on came from the black hole candidate Swift J1658.2--4242 \cite{Xiao2}, constituting one of the first targets for which multi-mission
observations including {\it Insight}-HXMT were set. These observations made precisions on the timing and spectral properties of the source, and promoted
that Swift J1658.2--4242 is indeed a black hole binary system.

{\it Insight}-HXMT also observed
MAXI J1535-571, discovered during its outburst in 2017, performing a joint spectral analysis (2--150 keV) in both energy and time \cite{Kong}. The energy spectra provided input for probing the intrinsic QPO fractional rms spectra (FRS): the QPO FRS become harder and promotes the possibility of having an additional power-law component. Is this indicative of the launch of a jet during the intermediate state of the outburst?

Among the key programmes of {\it Insight}-HXMT, it will realize a systematic Galactic plane scanning survey.
The methodology and performance of the survey is also analyzed in a paper of the collection \cite{Sai}.
{\it Insight}-HXMT performed more than 1000 scanning observations and the whole Galactic plane has been covered multiple times so far. More than 800 X-ray sources of different classes were monitored by the three telescopes onboard {\it Insight}-HXMT, for which long-term light curves in three energy bands covering 1-250 keV are available. This is a reach dataset, the full study of which will take years to complete.
The survey power is enhanced by means of a direct demodulation method for the detection of fast variable objects.
The method is based on a reconstruction of a modified response matrix, with a more proper handling of variability.
The overall result is
a better localization and flux monitoring.
The first application of the method revealed
type-II bursts
for the Rapid Burster MXB 1730--335, with an average rate of ~2 bursts min-1  compatible with relaxation oscillation.

Finally, {\it Insight}-HXMT also {\it looked near}. The South Atlantic Anomaly (SAA) is a region of weakened geomagnetic field showing a secular variation with time associated with the change of the magnetic moment.
This region was observed with the particle monitors onboard {\it Insight}-HXMT, characterizing its position, size, and shape, and how they evolve with time \cite{Zhao}.

Join us in celebrating these first results of the mission, and in forecasting
even more exciting results with {\it Insight}-HXMT and beyond.

\acknowledgments{The authors are editors of the Journal of High Energy Astrophysics. DFT edited the referred special collection for HXMT.
SNZ is the PI of HXMT mission.
All papers are available via astro-ph, ADS, and the journal website.}

\end{document}